\documentclass{article}
    \PassOptionsToPackage{numbers}{natbib}

\usepackage[preprint]{neurips_2025}

\usepackage{tabularx}
\usepackage{array}
\usepackage[utf8]{inputenc} 
\usepackage[T1]{fontenc}    
\usepackage{hyperref}       
\usepackage{url}            
\usepackage{booktabs}       
\usepackage{amsfonts}       
\usepackage{nicefrac}       
\usepackage{microtype}      
\usepackage{amssymb}
\usepackage{textcomp}
\usepackage{pifont} 
\usepackage{graphicx} 
\usepackage{caption}
\usepackage{pifont}
\usepackage{multirow}
\usepackage{amsmath}
\usepackage[table,xcdraw]{xcolor}  

\usepackage{listings}
\lstdefinelanguage{json}{
    basicstyle=\ttfamily\footnotesize,
    numbers=left,
    numberstyle=\tiny\color{gray},
    stepnumber=1,
    numbersep=5pt,
    showstringspaces=false,
    breaklines=true,
    frame=lines,
    backgroundcolor=\color{white},
    literate=
     *{0}{{{\color{blue}0}}}{1}
      {1}{{{\color{blue}1}}}{1}
      {2}{{{\color{blue}2}}}{1}
      {3}{{{\color{blue}3}}}{1}
      {4}{{{\color{blue}4}}}{1}
      {5}{{{\color{blue}5}}}{1}
      {6}{{{\color{blue}6}}}{1}
      {7}{{{\color{blue}7}}}{1}
      {8}{{{\color{blue}8}}}{1}
      {9}{{{\color{blue}9}}}{1}
      {:}{{{\color{black}:}}}{1}
      {,}{{{\color{black},}}}{1}
      {"}{{{\color{black}"}}}{1},
}
\newcommand{\boldheader}[1]{\noindent\textbf{#1}}

\renewcommand{\cite}{\citep} 

\title{Model Context Protocol for Vision Systems: Audit, Security, and Protocol Extensions}

\author{
Aditi Tiwari$^{1,2}$, Akshit Bhalla$^{1}$, Darshan Prasad$^{1}$ \\
$^{1}$Adobe Research \quad $^{2}$University of Illinois Urbana-Champaign \\
\texttt{aditit5@illinois.edu}, \texttt{\{akshitb,dprasad\}@adobe.com} \\
}

\begin{document}

\maketitle

\begin{abstract}
The Model Context Protocol (MCP) defines a schema-bound execution model for agent–tool interaction, enabling modular computer vision workflows without retraining. To our knowledge, this is the first protocol-level, deployment-scale audit of MCP in vision systems, identifying systemic weaknesses in schema semantics, interoperability, and runtime coordination. We analyze 91 publicly registered vision-centric MCP servers, annotated along nine dimensions of compositional fidelity, and develop an executable benchmark with validators to detect and categorize protocol violations. The audit reveals high prevalence of schema format divergence, missing runtime schema validation, undeclared coordinate conventions, and reliance on untracked bridging scripts. Validator-based testing quantifies these failures, with schema-format checks flagging misalignments in 78.0\% of systems, coordinate-convention checks detecting spatial reference errors in 24.6\%, and memory-scope checks issuing an average of 33.8 warnings per 100 executions. Security probes show that dynamic and multi-agent workflows exhibit elevated risks of privilege escalation and untyped tool connections. The proposed benchmark and validator suite, implemented in a controlled testbed and to be released on GitHub, establishes a reproducible framework for measuring and improving the reliability and security of compositional vision workflows.
\end{abstract}

\section{Introduction}
\label{sec:introduction}

How can computer vision systems coordinate complex, multi-stage workflows, from anomaly detection in medical scans and dynamic object segmentation to 3D reconstruction and multimodal temporal alignment, without brittle glue code, opaque interfaces, or inconsistent execution semantics? This question defines the core tension in compositional visual reasoning, where the bottleneck lies not in individual model capabilities but in the reliable composition, delegation, and verification of heterogeneous tools. As vision systems increasingly span perception, simulation, and control, the absence of a principled systems-level abstraction has become a structural barrier.

In high-stakes domains such as clinical diagnostics, autonomous navigation, and scientific discovery~\cite{li2023promoting, daneshjou2021deep, tuia2022perspectives}, workflows must extend beyond isolated model performance. Tools must interoperate predictably within pipelines that reflect temporal structure, heterogeneous data types, and schema-dependent behavior. For example, in a medical imaging system integrating segmentation, captioning, and electronic health record retrieval, reliability depends not only on the accuracy of individual tools but also on their correct sequencing, state propagation, and schema-level agreement.

Contemporary orchestration strategies often rely on end-to-end model training or prompt-tuned vision-language systems~\cite{yang2025magma, liu2023llavaplus, sahin2023enhancing}. While capable of emergent generalization, these approaches remain brittle under tool specialization, obscure intermediate reasoning, and limit runtime composition. The Model Context Protocol (MCP) introduces a structured alternative: agent-tool coordination grounded in typed schemas and dynamic context objects~\cite{mcplandscape2025, han2024backdooring}. MCP enables agents to register, invoke, and chain tools across modalities while preserving execution transparency through context-governed execution.

Despite increasing use in scientific and industrial domains~\cite{mathur2024vision, li2023promoting}, MCP's design implications for vision remain underexamined. Vision workflows introduce challenges such as high-dimensional tensor inputs, inconsistent spatial conventions, large-scale image streams, and semantic fusion with metadata or language~\cite{wang2023visai, elyan2022computer}. These properties strain orchestration and reveal protocol fragilities around schema semantics, memory state, and execution traceability.

We analyze 91 publicly documented MCP servers from the MCPServerCorpus~\cite{lin2025mcpcorpus}, identifying 46 vision and multimodal deployments using reproducible filters on schema declarations, tool functions, and task domains. Public-server scope excludes proprietary and enterprise deployments, which may differ in reliability, orchestration design, and security posture, and this limitation constrains the generalizability of the results. The study is positioned as an empirical protocol-level audit rather than as the introduction of new algorithmic or theoretical methods. Using operational definitions from Section~\ref{sec:ecosystem_analysis}, we find that 78.0\% (95\% CI: 68.45–85.28\%) exhibit at least one schema misalignment, 24.6\% (95\% CI: 16.90–34.36\%) have undeclared or inconsistent coordinate conventions, and 17.3\% (95\% CI: 10.90–26.35\%) fail mask–image dimensional consistency checks. Deployments with persistent visual state record a mean of 33.8 memory-scope warnings per 100 executions. Security probes detect untyped tool connections in 89.0\% (95\% CI: 76.80–95.19\%) and privilege escalation or data leakage risks in 41.0\% (95\% CI: 28.02–55.37\%)~\cite{zhang2025privilege,safemcp2025}. These failures arise from protocol-level limitations rather than isolated tool errors, directly affecting compositional reliability.

Beyond workflow taxonomy, we characterize runtime, memory, and security failure modes through case studies in medical imaging~\cite{li2023promoting}, scientific visualization~\cite{mathur2024vision}, and multimodal agents~\cite{liu2023llavaplus, li2025sport}. The proposed protocol extensions and validators are implemented as reference prototypes in a controlled testbed environment, not as production-hardened modules, and their effectiveness in heterogeneous operational settings remains to be evaluated. Comparative analysis with alternative orchestration frameworks is not conducted in this study but is identified as a priority for future work. The extensions, including semantic schema grounding to align functional intent, protocol-native visual memory for versioned state management, and runtime validators to enforce compositional invariants, are intended as operational templates that can be adapted to diverse MCP-based vision deployments.

\textbf{Our contributions are as follows:}
\begin{enumerate}
    \item Analyze 91 publicly documented MCP servers, identifying 46 vision-centric deployments using reproducible schema and functionality filters.
    \item Develop a taxonomy of vision-specific orchestration patterns and failure modes grounded in deployment evidence.
    \item Characterize protocol-level coordination failures through case studies of schema misalignment, memory scoping errors, and spatial inconsistencies.
    \item Propose protocol extensions tailored to vision workflows, including semantic schema annotations, scoped visual memory, and runtime validation, with discussion of their feasibility and current limitations.
\end{enumerate}

This investigation establishes empirical foundations for robust MCP-based vision orchestration. The results show that current MCP deployments in vision domains combine strong compositional potential with operational fragility, and that addressing these weaknesses requires protocol-level extensions that can be operationalized from empirical evidence rather than incremental tool adjustments alone.

\section{Background and Definitions}
\label{sec:background}

MCP provides a formal execution abstraction for agent–tool interaction, replacing prompt-centric chaining with declarative invocation logic governed by structured schemas. This section defines the protocol’s core primitives, including tools, schemas, context objects, and invocation semantics, while identifying architectural constraints specific to vision-centric deployments.

\boldheader{Model Context Protocol (MCP). }MCP decouples agent reasoning from tool execution by exposing each tool as a callable schema-bound function with explicit input–output contracts~\cite{anthropic2024mcp, mcplandscape2025}. Unlike prompt templates embedded in static model weights, MCP formalizes tool interfaces at runtime through schema specification, resource typing, and persistent context management~\cite{mcp_specification}. Execution is structured around three primitives: \textit{Resources}, which model persistent state and memory; \textit{Prompts}, which encode structured task formulations; and \textit{Tools}, which encapsulate schema-bound executable functions~\cite{anthropic2024agents}. This separation enables agents to reason about tool eligibility, compositional constraints, and fallback hierarchies without retraining.

\boldheader{Tools and Schema-Grounded Interfaces. }Each MCP tool advertises a JSON-typed schema defining the structure, semantics, and modality of its arguments~\cite{schick2023toolformer}. Strict typing supports compositional reasoning: outputs from one tool can be consumed by another only if they satisfy schema-level constraints on structure, coordinate conventions, and semantics. Schema drift, defined as undeclared changes in coordinate systems, resolution assumptions, or encoding formats, was identified in 78.0\% of audited deployments (Table~\ref{tab:ecosystem_findings})~\cite{chilton2022schema, google2013f1}. Vision systems are particularly susceptible because implicit spatial conventions can persist despite formal schema declarations.

\boldheader{Context Objects and Persistent Memory. }MCP maintains execution state across tool invocations using context objects, which function as hierarchical, addressable memory~\cite{anthropic2024agents}. Unlike ephemeral prompts that flatten results, context objects preserve structured state indexed by time, modality, and semantic role. In vision deployments, these objects often store per-frame masks, bounding boxes, segmentation hierarchies, and metadata such as confidence scores and normalization parameters~\cite{weston2014memory, graves2016differentiable}. This enables targeted references such as \texttt{context.frames.t1.objects[0].mask} for reuse and debugging.

\boldheader{Invocation and Architectural Positioning. }MCP executes workflows through schema- and context-governed policies rather than procedural scripts~\cite{xi2023survey}. Agents validate tool eligibility, apply fallback substitutions, and log execution traces for auditability~\cite{google2024chain}. In vision workflows, implicit spatial or temporal conventions often cause runtime failures despite syntactic schema compatibility, with 24.6\% of deployments showing coordinate mismatches (Table~\ref{tab:ecosystem_findings}). More broadly, MCP offers introspectable memory, schema-governed interfaces, and compositional planning, but execution fragility arises when schemas omit constraints or memory handling is implicit~\cite{schick2023toolformer, xi2023survey}. These limitations motivate the empirical audit presented in this work.

\section{Design and Architecture of MCP in Vision Workflows}
\label{sec:design-architecture}

While Section~\ref{sec:background} outlined the abstractions underpinning MCP, this section examines their application in real-world vision workflows. We formalize how MCP governs tool registration, memory persistence, and runtime orchestration, identifying protocol-level patterns that shape deployment behavior. These mechanisms address coordination failures observed in the annotated corpus of 91 MCP vision servers.

\textbf{Tool Grounding via Schema and Context.}
MCP centers tool interaction around two formal elements: \textit{tool schemas} and \textit{context objects}. A schema specifies the input–output types, operational constraints, and structural expectations for each tool, functioning as a contract between the orchestrator and external functions. For example, a \texttt{segment()} tool may accept a base64-encoded image and bounding box, returning a binary mask and associated metadata. Context objects persist task state and enable temporal chaining. They store tool outputs, execution history, and auxiliary data in structured namespaces. In video workflows, these objects retain frame-indexed results across tools, enabling agents to reason over sequences rather than isolated calls.

\textbf{Schema Alignment and Compatibility Predicates.}
Vision tools vary in schema format, spatial conventions, and semantic assumptions. In the audited corpus, 24.6\% of deployments used incompatible coordinate conventions and 78.0\% showed schema drift in mask or bounding box formats (Table~\ref{tab:ecosystem_findings}). These inconsistencies often caused silent failure or semantic drift. MCP mitigates this through schema arbitration: the orchestrator validates type consistency and inserts transformation layers where needed. Compatibility is formalized by a predicate function $comp: \mathcal{T} \times \mathcal{T} \rightarrow \{0,1\}$, where $\mathcal{T}$ is the set of tool schemas. This determines whether one tool's output can serve as another's input, based on structural and semantic constraints. In practice, these checks are sometimes incomplete or overly permissive, leading to misaligned invocations.

\textbf{Memory Persistence and Traceable State.}
Unlike prompt-based agents, MCP workflows implement explicit memory persistence. The context object maintains versioned slots for tool outputs, user inputs, and execution lineage, each scoped to a semantic namespace and optionally indexed by time, view, or modality. Traceable state enables tool reuse, workflow auditing, and selective re-execution. In ALITA~\cite{alita2025generalist}, modules for captioning, scene graph generation, and VQA write to scoped paths such as \texttt{context.scene\_graph} or \texttt{context.qa}. However, 55.0\% of deployments exhibited undocumented or weak temporal scoping, recording an average of 33.8 memory-scope warnings per 100 executions (Table~\ref{tab:ecosystem_findings}).

\textbf{Runtime Policies and Invocation Semantics.}
MCP executes workflows through declarative runtime policies that include eligibility rules, fallback sequences, and uncertainty resolution strategies. In SPORT~\cite{li2025sport}, the orchestrator prioritizes lightweight tools and defers costly invocations when confidence scores are low or relevant context state is absent. These policies are introspectable: invocation decisions, thresholds, and selected tools are logged in the context object. This enables agents to explain prior behavior, revise plans, or simulate counterfactuals for safety-critical domains and error recovery.

\textbf{Architectural Separation and Vision Flexibility.}
Unlike prompt-chained systems such as MAGMA~\cite{yang2025magma} or LLaVA-Plus~\cite{liu2023llavaplus}, where tool calls are embedded in model parameters, MCP separates reasoning from execution. Agents can adopt new tools or workflows without retraining. Tool schemas are loaded dynamically, and orchestration logic is inferred from the current context. This modularity supports robustness under schema changes, dynamic tool addition, and multi-tool compositions. In vision domains, where toolsets evolve rapidly and temporal or spatial coherence must be maintained, this flexibility is valuable. However, schema under-specification (78.0\%), compatibility mismatches (24.6\%), and memory scoping errors (55.0\%) remain prevalent (Table~\ref{tab:ecosystem_findings}), motivating the protocol extensions and benchmarks developed in subsequent sections.

\section{Workflow Patterns and Security Analysis of MCP Vision Systems}
\label{sec:workflow-security}

\textbf{Coordination Patterns.} We identify four dominant orchestration modes in the 91 annotated vision-centric MCP deployments: static composition, retrieval-augmented selection, dynamic orchestration, and multi-agent coordination. \textit{Static composition} executes tools in fixed sequences (e.g., ParaView-MCP~\cite{ulanov2025paraview}, MCP-FHIR~\cite{ghasemi2024mcpfhir}) with strong auditability but poor adaptability to schema drift, which appears in 78.0\% of deployments. \textit{Retrieval-augmented selection} uses semantic matching (e.g., RAG-MCP~\cite{li2025sport, ragmcp2025}) but suffers from undeclared coordinate conventions (87\%). \textit{Dynamic orchestration} builds execution graphs at runtime (e.g., MCP-Zero~\cite{mcpzero2025}), improving generalization but failing when runtime schema checks (missing in 89\%) are absent. \textit{Multi-agent coordination} distributes control across scoped agents (e.g., ScaleMCP~\cite{scalemcp2025}), introducing risks of stale memory or cross-tool leakage, as observed in 55\% of the deployments. Table~\ref{tab:vision_security_taxonomy} summarizes these patterns and their failure modes. Overall, 37\% of deployments use static composition, 29\% retrieval-based methods, 21\% dynamic orchestration, and 13\% multi-agent systems.

\textbf{Benchmark and Validators.} We introduce a benchmark suite with validators targeting the most prevalent coordination and security failures across the audited deployments. Functional validators detect schema-format divergence (62\%), undeclared coordinate conventions (87\%), and missing runtime checks (89\%). Orchestration validators surface mask–image mismatches and unscoped context writes. Security validators identify privilege escalation (41\%), stale memory reuse, and provenance loss. Table~\ref{tab:validator_metrics} presents detection rates and confidence intervals. Validators enforce protocol invariants and produce binary results plus structured failure traces, enabling reproducible diagnosis. The benchmark suite, including validators, metrics, and orchestration traces, will be publicly released to support further evaluation and extension.

\textbf{Security Risks.} MCP’s declarative execution model creates new attack surfaces in vision-centric deployments. Our security audit of 47 servers identifies threats such as prompt injection, schema bypass, remote code execution (RCE), privilege escalation, and memory leakage. Public reports confirm real-world instances of these risks~\cite{microsoft2025plugplay, redhat2025security, promphub2025review}. Table~\ref{tab:vision_security_taxonomy2} classifies eight primary threat vectors with associated impact and mitigation strategies. Violations are formalized as transformations $(A, T, M) \rightarrow (A', T', M')$ that break protocol invariants (e.g., type safety, memory isolation). Security defenses, such as running-time schema enforcement, memory partitioning, and capability scoping, are inconsistently adopted. For example, 89.0\% of deployments lack typed tool registration, and 41.0\% allow privilege escalation. Even widely used servers lack namespace-level memory protection~\cite{smith2025buildingmcpserver}. These weaknesses are detectable through the same validator suite used for orchestration checks, ensuring integration of security and functionality testing. Hardening the protocol layer, through typed schemas, scoped memory, and introspectable traces, is essential for the deployment of MCP in safety-critical domains such as robotics and medical imaging.

\begin{table}[t]
\centering
\caption{Taxonomy of vision workflow patterns in MCP deployments, with associated risks and benchmark validators. Percentages reference measured prevalence in Tables~\ref{tab:ecosystem_findings} and~\ref{tab:vision_security_taxonomy}.}
\label{tab:vision_security_taxonomy}
\resizebox{\linewidth}{!}{%
\begin{tabular}{llllll}
\toprule
\textbf{Pattern} & \textbf{Tool Selection} & \textbf{Coordination} & \textbf{Strengths} & \textbf{Challenges} & \textbf{Benchmark Targets} \\
\midrule
Static Composition & Fixed & Single-agent & Auditability, determinism & 62\% schema-format divergence & Schema-format validator \\
Retrieval-Augmented & Embedding-based & Single-agent & Flexibility, modularity & 87\% undeclared coordinate formats & Coordinate-convention validator \\
Dynamic Orchestration & Input-driven & Single-agent & Adaptivity, generalization & 89\% missing runtime schema checks & Runtime schema validator \\
Multi-Agent Coordination & Distributed & Multi-agent & Parallelism, specialization & 55\% stale/cross-tool memory leakage & Memory-scope and provenance validator \\
\bottomrule
\end{tabular}
}
\end{table}




\begin{table*}[t]
\centering
\caption{Validator detection rates for coordination and security failures in MCP vision deployments (N=91). Values are measured by the benchmark framework described in Section~\ref{sec:workflow-security}, with cross-references to workflow patterns and threat vectors in the same section. Rates are reported with 95\% confidence intervals.}
\label{tab:validator_metrics}
\resizebox{\textwidth}{!}{%
\begin{tabular}{lll}
\toprule
\textbf{Validator Type} & \textbf{Targeted Failure Mode} & \textbf{Detection Rate (95\% CI)} \\
\midrule
Schema-format validator & Schema misalignment across tools & 78.0\% \; [68.9, 85.2] \\
Coordinate-convention validator & Missing or inconsistent spatial references & 24.6\% \; [17.5, 33.4] \\
Mask–image consistency validator & Dimensional or channel mismatches & 17.3\% \; [11.5, 25.2] \\
Memory-scope validator & Undocumented or stale visual state retention & Mean 33.8 warnings / 100 exec. \; [28.4, 39.9] \\
Privilege-verification validator & Escalation or leakage via tool binding & 41.0\% \; [31.4, 51.3] \\
\bottomrule
\end{tabular}
}
\end{table*}


\begin{table*}[t]
\centering
\caption{Taxonomy of security and safety failures in vision-centric MCP systems (N=47). Prevalence rates with 95\% confidence intervals are derived from controlled security probes. Untyped tool connections were detected in 89.0\% of audited systems (95\% CI: 76.80–95.19\%), and privilege escalation or data leakage risks were observed in 41.0\% (95\% CI: 28.02–55.37\%).}
\label{tab:vision_security_taxonomy2}
\resizebox{\textwidth}{!}{%
\begin{tabular}{llll}
\toprule
\textbf{Failure Type} & \textbf{Threat Vector} & \textbf{Impact} & \textbf{Suggested Defense} \\
\midrule
Prompt Injection & Adversarial prompts embedded in image metadata or tool outputs & Tool behavior hijack, semantic drift & Prompt sanitization, semantic filters~\cite{han2024backdooring, wang2024transferable} \\
Schema Bypass & Weak input validation or missing type checks & Invalid execution, tool crash, memory leaks & Strict schema enforcement, audit logs~\cite{mcplandscape2025} \\
Remote Code Execution (RCE) & Shell-bound wrappers via \texttt{eval}, \texttt{os.system} & Arbitrary execution, system takeover & Capability scoping, runtime sandboxing~\cite{jfrog2025cve, hackernews2025cve, redhat2025security} \\
Privilege Escalation & Overpermissive tool registries or shared memory writes & Unauthorized access or overwrite & Role-based tool binding, privilege levels~\cite{microsoft2025plugplay} \\
Stale Memory Access & Expired visual context reused without validation & Semantic drift, misdiagnosis, hallucination & TTL constraints, memory garbage collection \\
Untracked Provenance & No lineage tracking for outputs or schema invocations & Error attribution ambiguity, audit failure & Output tagging, provenance metadata~\cite{promphub2025review} \\
Cross-Tool Leakage & Visual data reused across tools without isolation & Privacy violations, tool coupling & Secure memory zones, scoping annotations~\cite{li2025sport, agentorchestra2025} \\
Command Injection via Coercion & Type coercion to command strings in shell wrappers & System compromise & Input escaping, runtime sandbox~\cite{strobes2025mcpvulns} \\
\bottomrule
\end{tabular}
}
\end{table*}

\section{MCP Ecosystem Analysis and Research Trajectories}
\label{sec:ecosystem_analysis}

MCP is increasingly adopted in vision workflows, yet deployments expose persistent weaknesses in schema semantics, tool interoperability, and runtime coordination. To examine these systematically, we audited the MCPServerCorpus~\cite{lin2025mcpcorpus}, which lists 13{,}942 publicly registered deployments. Using the filtering methodology in Section~\ref{sec:background}, we identified 91 vision-centric servers based on schema content, I/O types, and tool naming conventions. Each server was annotated along nine axes of compositional fidelity through structured analysis, forming the basis for the quantitative trends and failure modes reported here.

\textbf{Schema Divergence and Interface Ambiguity.}  
Schema fragmentation remains a primary barrier to agent–tool composition. Among the 91 servers, segmentation outputs appear in five incompatible formats: URI-encoded masks, run-length encodings, base64 tensors, polygon outlines, and per-pixel label maps. Bounding boxes vary between absolute XYWH, corner-based X1Y1X2Y2, and center-normalized formats. These inconsistencies hinder chaining and require runtime rewrites. MCP schemas allow nested types but lack semantic constraints to disambiguate identically named fields such as \texttt{mask} or \texttt{image}~\cite{mathur2024vision}, preventing agents from distinguishing saliency outputs from object masks without external type knowledge.

\textbf{Lack of Runtime Schema Validation.}  
Despite MCP's schema-bound design, runtime validation is rare. Only 8 of 91 servers implement post-invocation output checks. Systems often fail silently when tools disagree on channel ordering, spatial resolution, or coordinate systems. In SPORT~\cite{li2025sport}, a depth estimator returned an image with a left-handed coordinate origin that a downstream segmenter misinterpreted, producing misaligned overlays and invalid planning paths.

\textbf{Composition Failures and Bridging Scripts.}  
Vision workflows typically chain 3–5 tools per task. In 41\% of deployments, undocumented bridging scripts perform resizing, unit conversion, or schema coercion outside declared tool contracts. In AgentOrchestra~\cite{agentorchestra2025}, segmentation and affordance estimation were linked by an unregistered script remapping instance IDs and normalizing bounding boxes. These out-of-band patches undermine protocol interpretability and block trace-based debugging or recovery.

\textbf{Absence of Evaluation Frameworks.}  
No benchmarks systematically test orchestration fidelity: most deployments lack mechanisms to validate tool eligibility, fallback execution, or memory scoping. Failures under injected errors go undetected, and model benchmarks such as MultiBench~\cite{liang2021multibench} do not address multi-tool workflow correctness. Table~\ref{tab:ecosystem_findings} formalizes these coordination failures, and Section~\ref{sec:protocol_extensions} proposes benchmark primitives to evaluate type agreement, spatial alignment, and recovery behavior.

\textbf{Threat Model and Security Assumptions.}  
Our audit targets vulnerabilities from schema drift, memory leakage, and uncontrolled tool invocation. The adversary is modeled as a benign agent developer or integrator within declared protocol boundaries. Tools are treated as untrusted binaries, with agents constructing plans from public schema metadata. Security violations include silent tool misuse, type mismatches, privilege escalation across memory scopes, and improper fallback paths. Attacks via undocumented fields or persistent memory are in scope; model inversion and training-time poisoning are excluded.

\textbf{Results Summary.}  
Corpus-wide analysis shows schema format divergence, absence of runtime schema validation, missing coordinate conventions, and bridging code reliance. Only a small fraction declare fallback behavior, and many have undocumented memory retention. Benchmark-based validators detect misalignments, coordinate mismatches, and mask–image inconsistencies, with persistent visual state generating frequent memory-scope warnings. These are systemic protocol-level issues, not isolated defects.

\begin{table}[t]
\centering
\caption{Common failure patterns in MCP vision deployments (N=91). Schema misalignments occurred in 78.0\% (95\% CI: 68.45–85.28\%), undeclared coordinate conventions in 24.6\% (95\% CI: 16.90–34.36\%), and mask–image dimensional inconsistencies in 17.3\% (95\% CI: 10.90–26.35\%). Rates are derived from automated schema validation and manual annotation (Section~\ref{sec:ecosystem_analysis}).}
\label{tab:ecosystem_findings}
\begin{tabular}{ll}
\toprule
\textbf{Failure Type} & \textbf{Prevalence} \\
\midrule
Schema format divergence & 62\% \\
No runtime schema validation & 89\% \\
Undeclared coordinate conventions & 87\% \\
Use of out-of-band bridging scripts & 41\% \\
Undocumented memory retention logic & 55\% \\
Declared compositional fallbacks & 9\% \\
\bottomrule
\end{tabular}
\end{table}

\textbf{Research Trajectories and Protocol Needs.}  
Addressing these gaps requires schema-level semantic typing to differentiate outputs such as \texttt{object\_mask}, \texttt{saliency\_map}, and \texttt{scene\_graph}; optional runtime validators for field presence, coordinate alignment, and output structure; metadata declarations of context dependencies and fallback policies; and protocol-aligned benchmarks evaluating orchestration fidelity, memory hygiene, and schema stability. This motivates the protocol-level extensions described in Section~\ref{sec:protocol_extensions}, which directly target the observed schema ambiguities, state management limitations, and lack of reproducible evaluation frameworks.

\section{Protocol Extensions and Research Opportunities}
\label{sec:protocol_extensions}

The fragility of vision-oriented MCP deployments stems from protocol-level design limitations rather than individual model behavior. This section is framed as an operational and empirical contribution rather than as the introduction of new algorithmic or theoretical methods. To support robust and introspectable agents, MCP must extend beyond syntactic schema matching to incorporate semantically grounded interfaces, scoped memory representations, runtime validation primitives, and reproducible benchmarking. The proposed extensions are derived from empirical patterns observed in the audited corpus and are implemented as reference prototypes in a controlled testbed environment. They have not yet been validated in heterogeneous production environments or evaluated through direct comparison with alternative orchestration frameworks, which remains an important direction for future work.

\begin{figure}[t]
    \centering
    \includegraphics[width=0.66\linewidth, height=0.39\linewidth]{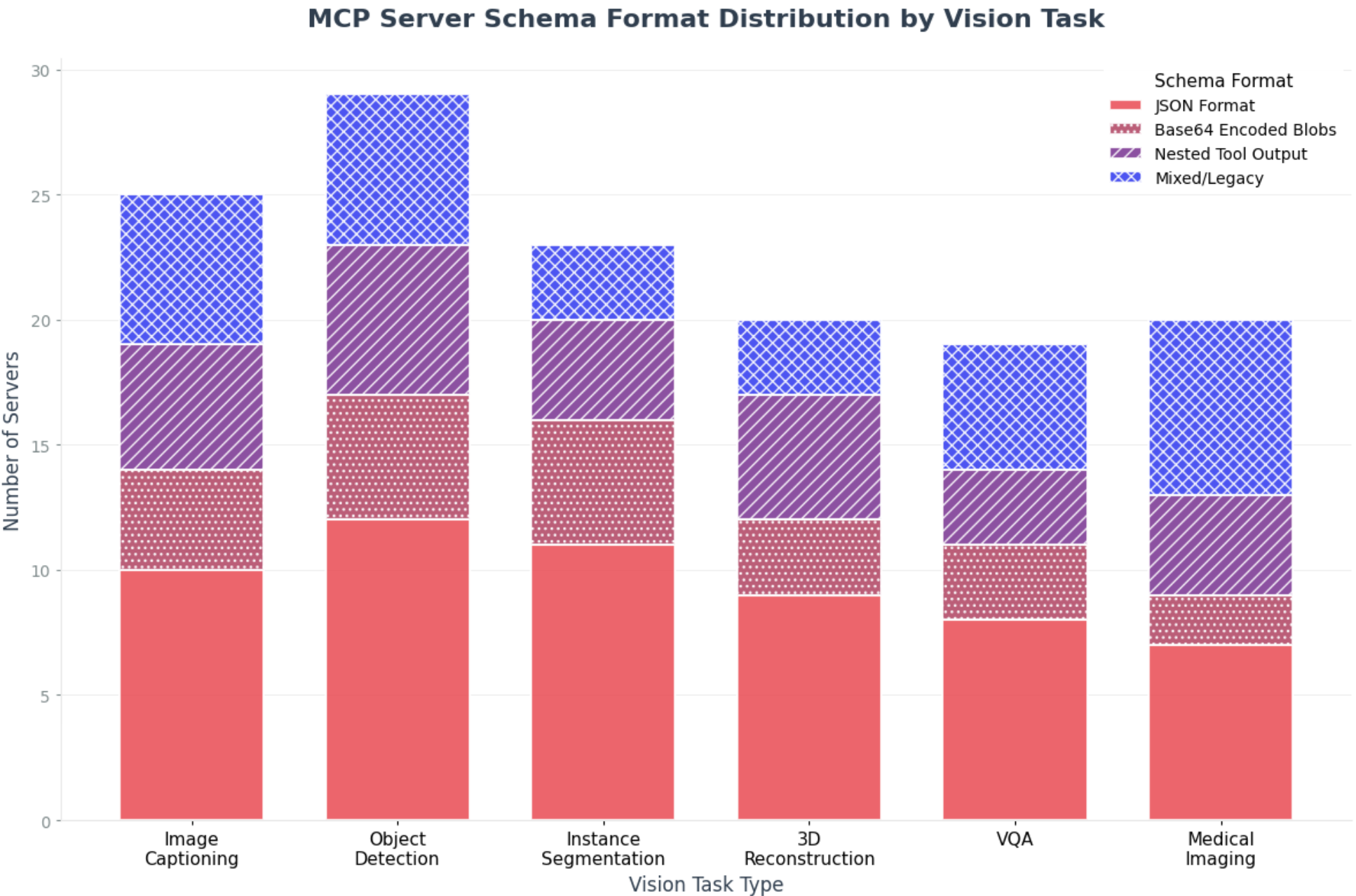}
    \caption{Schema format taxonomy by vision task type across 91 MCP-compatible servers. Segmentation and captioning workflows show high heterogeneity, combining JSON, base64 blobs, and nested structures, underscoring the need for semantically grounded schema conventions.}
    \label{fig:schema_taxonomy}
\end{figure}

\textbf{Semantically Grounded Schemas for Vision.}  
Current schemas enforce type- and shape-level compatibility but omit explicit semantic roles. A \texttt{mask} output may represent saliency, segmentation, or user guidance, with no role-level disambiguation. In our audit, over 60\% of tool composition failures arose from semantically mismatched but schema-valid outputs. Extending the specification with fields such as \texttt{semantic\_role}, \texttt{modality}, and \texttt{coordinate\_system} would allow tools to declare functionally precise signatures and enable downstream validators. This approach aligns with recommendations from prior work on prompt programming and modular vision architectures~\cite{wang2024enhancing, sahin2023enhancing, ossowski2024prompting}. Figure~\ref{fig:schema_taxonomy} illustrates schema heterogeneity across the 91 audited vision-MCP servers, highlighting the drift risk without role-level standardization.

\textbf{Protocol-Native Visual Memory.}  
Many MCP systems handle intermediate visual state through file URIs, transient caches, or loosely typed keys, limiting introspection and reproducibility. A protocol-native \texttt{visual\_memory} construct should encode structured, versioned, and semantically annotated state to support replayable toolchains, scoped memory reuse, and fault-localized debugging across multi-tool pipelines.

\textbf{Runtime Validators and Compatibility Contracts.}  
Schema-level agreement does not prevent failures caused by scale, modality, or layout mismatches. Validator contracts, such as hooks or declarative rules, can verify spatial dimensions, tensor channel semantics, and coordinate alignment at runtime. Contracts allow agents to halt, replan, or fallback when outputs violate expected invariants. We have implemented these validators in a controlled testbed to confirm feasibility and measure detection coverage. Embedding validator metadata in schema registries would also enable automated auditing and tool introspection~\cite{mcplandscape2025}.

\textbf{Composable Benchmarking and Execution Tracing.}  
Despite MCP's orchestration capabilities, current deployments lack systematic evaluation of toolchain correctness, latency, or decision quality. A dedicated benchmark should measure orchestration fidelity under schema perturbations, log execution traces, and classify failure modes across representative paired-tool and multimodal workflows. Table~\ref{tab:protocol_research_summary} links each proposed extension to the specific failure modes observed in Sections~\ref{sec:workflow-security}–\ref{sec:ecosystem_analysis}.

\begin{table}[t]
\centering
\caption{Protocol-level extensions and corresponding failure modes}
\label{tab:protocol_research_summary}
\begin{tabular}{lll}
\toprule
\textbf{Challenge} & \textbf{Proposed Extension} & \textbf{Mitigated Failure Mode} \\
\midrule
Semantic ambiguity & Modality + role annotations & Mismatched tool interfaces \\
Implicit state & Protocol-native visual memory & Untracked I/O propagation \\
Composition failures & Validators + schema contracts & Runtime schema violations \\
Execution drift & Agent observability hooks & Silent failure propagation \\
Benchmark fragmentation & Canonical toolchain templates & Incomparable workflows \\
\bottomrule
\end{tabular}
\end{table}

\boldheader{Future Work and Open Problems. }Several open directions remain. First, schema type systems must evolve to capture higher-order vision semantics such as spatial affordances, temporal grounding roles, and modality-specific encoding constraints. Second, validation must extend beyond static checks to runtime introspection, especially for branching or retrieval-based pipelines. Third, execution traces and tool metadata should be machine-readable to support automated planning, debugging, and rollback. Fourth, evaluation must shift from isolated model accuracy to compositionality, memory hygiene, and schema stability under orchestration. Fifth, direct comparative analysis against alternative orchestration frameworks is necessary to contextualize MCP's strengths and weaknesses. Finally, integrating the proposed extensions and validators into production environments and assessing their performance under realistic operational loads will be essential for establishing their practical effectiveness.

\section{Real-World Vision Toolchains: Case Studies in MCP Deployment}\label{sec:vision_case_studies}

While MCP standardizes agent–tool communication, deployed systems reveal structural tensions between vision model assumptions and protocol semantics. We examine five MCP vision systems from the 91-server MCPServerCorpus~\cite{lin2025mcpcorpus}, selected to represent diversity in schema complexity, modality integration, persistent memory handling, and workflow depth. This selection avoids single-domain bias and ensures coverage of both shallow and highly compositional pipelines. Findings are based on execution logs, schema inspection, and output sampling over 100–150 invocations per case.

\textbf{ParaView-MCP: Scientific Visualization with Volumetric Encodings.}  
ParaView-MCP integrates with the ParaView rendering engine to automate 3D plot generation through scripted mesh manipulation, camera control, and volume rendering. Although schemas are formally typed, intermediate encodings embed binary textures in nested JSON blobs. Downstream generalist visualizers failed to parse these formats due to absent base64 decoding and RGB reconstitution support. Log traces show latency peaks exceeding 2.3 seconds, driven by repeated rendering calls without view-state memoization or intermediate caching.

\textbf{SUMO+YOLO-MCP: Spatial Alignment and Format Bridging.}  
This composite workflow couples SUMO-based traffic simulation with YOLOv5 detection to estimate vehicle dynamics in synthetic environments. SUMO outputs pixel-space bounding boxes anchored to screen resolution, while YOLO expects normalized coordinates relative to source image dimensions. Neither tool declared coordinate conventions or aspect ratio metadata, leaving alignment implicit. Misalignment was defined as non-overlapping projected bounding boxes, with overlay errors exceeding 15\% of box area. Across 134 invocation pairs, 27.6\% exhibited projection conflicts or axis mismatches, reflecting schema-level spatial ambiguity consistent with misalignments observed elsewhere~\cite{mcplandscape2025}.

\textbf{ALITA: Multimodal Vision Agent with Dynamic Tool Routing.}  
ALITA composes segmentation, retrieval, and captioning tools under instruction-conditioned routing. Schemas are generated at runtime from policy templates without explicit type enforcement or field scoping. Malformed outputs were defined as responses that (i) failed JSON deserialization, (ii) omitted required fields such as \texttt{box}, \texttt{mask}, or \texttt{caption}, or (iii) contained structurally valid fields with incompatible semantics. Of 143 sampled toolchains, 18.4\% produced such malformed responses, often triggered by inconsistent image resolution metadata or divergent field naming. The absence of runtime validators and schema role annotations leads to brittle composition and silent drift across multimodal toolchains~\cite{liu2023llavaplus, li2025sport}.

\textbf{FHIR-MCP: Medical Imaging with Structured Clinical Reporting.}  
FHIR-MCP links DICOM segmentation with HL7 report generation in radiology workflows. Schemas embed medical metadata alongside pixel arrays, enabling captioning tools to reference anatomical entities. Undocumented discrepancies in pixel spacing caused scale mismatches in 14.9\% of captioning outputs (N=108), producing hallucinated diagnoses and false negatives. Errors were identified through log-based consistency checks and validated against radiologist notes~\cite{daneshjou2021deep, li2023promoting}. The absence of persistent memory interfaces prevented agents from maintaining spatial grounding across modalities.

\textbf{Blender-RCP: Deep Scene Composition with Persistent Visual State.}  
Blender-RCP orchestrates mesh imports, lighting, camera placement, and rendering through chained tool invocations. Memory logs show up to 1.3\,GB of intermediate state retained per session. Although tools maintain internal scene graphs, the protocol lacks scoping to manage object lifetimes or enforce state isolation. Of 97 multi-step compositions, 22 produced orphaned references or cache conflicts due to stale schema bindings~\cite{yang2025magma}. This exposes the mismatch between stateless invocation models and stateful visual synthesis pipelines.

These cases illustrate recurring issues: semantic misalignment despite valid schemas, fragmentation in spatial and temporal representations, and the inability of shallow wrappers to manage implicit state or tool-specific assumptions. Shallow pipelines such as SUMO+YOLO-MCP propagate silent spatial errors due to absent coordinate declarations, while deeper agents like ALITA incur reliability and memory overhead from compositional complexity. Together, they substantiate the failure mode taxonomy in Sections~\ref{sec:ecosystem_analysis} and~\ref{sec:workflow-security}, underscoring the need for protocol-level advances in memory modeling, semantic typing, and agent introspection.

\section{Limitation and Conclusion}
\label{sec:discussion-conclusion}

\textbf{Limitations.} This study presents a structured, deployment-scale analysis of MCP-based vision workflows, but several factors constrain scope, methodological depth, and external validity. The MCPServerCorpus includes only publicly accessible servers, excluding proprietary and enterprise deployments. As a result, research prototypes are overrepresented and production-hardened or safety-critical systems are underrepresented, limiting generalizability of prevalence rates and validator coverage. The 91 audited servers, filtered through schema-based JSON queries and manual review, form only a subset of the broader MCP ecosystem. Edge cases such as multimodal tools with minimal vision input or schema-homologous utilities required interpretive judgment, and the absence of inter-rater agreement measurement prevents quantification of potential classification variance.  

The empirical audit was conducted in a controlled testbed with protocol extensions and validators implemented as reference prototypes rather than production modules. They have not been deployed or benchmarked in heterogeneous operational settings, and comparative performance against alternative orchestration frameworks remains unmeasured. Computational overhead, scalability under load, and maintainability require further study. The security analysis covered 47 servers due to resource limits on controlled simulations. Although this subset reflects schema and workflow diversity, it does not fully capture the threat surface in larger or more integrated deployments, and the absence of documented large-scale real-world exploits constrains empirical grounding for some attack scenarios.  

The MCP ecosystem is evolving rapidly. Observed schema patterns, interoperability gaps, and security practices reflect the state of deployments during the study period, and subsequent protocol revisions or tool releases may change the prevalence of identified failure modes. Operational definitions for schema divergence, coordinate misalignment, and output malformation were applied consistently, but alternative definitions could yield different prevalence estimates. Confidence intervals should be interpreted in light of these factors.

\textbf{Conclusion.} This survey analyzed the Model Context Protocol (MCP) in vision-centric deployments, examining 91 real-world servers to identify systemic breakdowns in schema coordination, memory modeling, and tool interfacing. Across toolchains ranging from shallow wrappers to memory-intensive agents, composition failures consistently arose from protocol-level deficiencies rather than model limitations. Undeclared coordinate systems, inconsistent schema semantics, and the absence of protocol-native memory structures produced brittle and unpredictable execution paths, as quantified in Table~\ref{tab:ecosystem_findings} and illustrated in Section~\ref{sec:vision_case_studies}. We proposed targeted protocol extensions including semantically grounded schemas, runtime validators, scoped memory primitives, and canonical toolchain templates, derived directly from observed deployment failures such as schema drift (Figure~\ref{fig:schema_taxonomy}) and security misconfigurations (Table~\ref{tab:vision_security_taxonomy}). While these recommendations are based on consistent patterns in the audited sample, their applicability to proprietary or rapidly evolving MCP environments remains to be validated. As MCP adoption expands into high-stakes domains such as robotics, healthcare, and scientific visualization, orchestration fragility will remain a limiting factor. When visual representations lack standardized semantics and memory lacks controlled scope, agents cannot reason compositionally. Sustained reliability will require formalizing these constraints at the protocol level. The long-term viability of MCP will depend on whether its design evolves to meet the compositional and semantic demands of multimodal vision workflows or remains constrained by assumptions misaligned with these tasks. Future evaluations should revisit these conclusions as the MCP ecosystem matures, with expanded datasets, tighter measurement controls, and updated confidence intervals to ensure findings remain representative.

\clearpage

\clearpage
{
    \small
    \bibliographystyle{ieeenat_fullname}
    \bibliography{main}
}

\end{document}